\documentclass[pdflatex,sn-mathphys-num]{sn-jnl}

\usepackage{graphicx}%
\usepackage{subfig}
\usepackage{multirow}%
\usepackage{amsmath,amssymb,amsfonts}%
\usepackage{amsthm}%
\usepackage{mathrsfs}%
\usepackage[title]{appendix}%
\usepackage{xcolor}%
\usepackage{textcomp}%
\usepackage{manyfoot}%
\usepackage{booktabs}%
\usepackage{algorithm}%
\usepackage{algorithmicx}%
\usepackage{algpseudocode}%
\usepackage{listings}%
\usepackage{comment}
\usepackage{enumerate}
\usepackage{booktabs} 
\usepackage{array}    

\newcommand{\X}{\mathbf{X}}

\newcommand{\Z}{\mathbf{Z}}

\newcommand{\W}{\mathbb{W}}
\newcommand{\I}{\mathbb{I}}

\begin{document}

\author*[1]{\fnm{Ranadeep} \sur{Daw}}\email{rdaw@uwf.edu}

\author[1]{\fnm{Hunter N.}  \sur{Evans}}\email{hne8@students.uwf.edu}

\author[2]{\fnm{Indrabati} \sur{Bhattacharya}}\email{ib22g@fsu.edu}

\affil*[1]{\orgdiv{Department of Mathematics and Statistics}, \orgname{University of West Florida}, \orgaddress{\street{11000 University Pkwy}, \city{Pensacola}, \postcode{32514}, \state{Florida}, \country{USA}}}


\affil[2]{\orgdiv{Department of Statistics}, \orgname{Florida State University}, \orgaddress{\street{117 N. Woodward Ave.}, \city{Tallahassee}, \postcode{32306}, \state{Florida}, \country{USA}}}

\title[Causal Analysis for Physical Morbidity]{Socioeconomic Drivers of Physical Morbidity Across U.S. Counties: A Spatial Causal Inference Approach}

\abstract{
Identifying the causal effects of socioeconomic determinants on population health is of many great interests -- from statistical methodology development to public health practitioners and policy developments. The statistical side of the problem needs to address several challenges: spatial autocorrelation in both exposures and outcomes, the treatment-confounder relationship, and the need for geographically logical inference. We address these jointly by using spectral basis functions -- Moran Eigenvector Maps and ICAR precision matrix eigenvectors -- within a doubly robust generalized propensity score estimator for continuous treatments. Applied to 2022 county health data across the U.S. counties, the framework identifies the effect of four chosen treatments on the average physically unhealthy days per 
month. Possible further applications and methodological extensions are also discussed as future directions from this research.
}

\keywords{
Physically unhealthy days, spatial statistics, basis functions, causal inference, doubly-robust estimator
}

\maketitle

\section{Introduction}
\label{sec:intro}

Population health outcomes in the United States exhibit striking geographic 
heterogeneity. Average physically and mentally unhealthy days, rates of chronic disease, and  access to care vary substantially across counties. These patterns reflect the uneven geographic distribution of socioeconomic resources, environmental exposures, and structural disadvantage \citep{braveman2011social, remington2015county} rather than chance. Understanding which of these factors drive poor health outcomes, and by how much, is essential for designing targeted public health interventions.

In this study, we focus on modeling average physically unhealthy days per month across U.S. counties  as a function of socioeconomic and environmental determinants under a causal analysis framework. Causal estimation in this setting presents two key challenges. First, both the health outcomes and their predictors show strong spatial autocorrelation, and neighboring counties tend to resemble each other due to shared infrastructure and unmeasured local conditions. Ignoring this structure induces \emph{spatial confounding} \textcolor{black}{as described by \citep{papadogeorgou2023spatial, gilbert2021causal} (see also \citep{ogunsola2026disentangling} for a clear distinction among the different uses of the term ``spatial confounding''),} which arises when latent geographic processes jointly influence both the treatment and the outcome \citep{clayton1993spatial, paciorek2010importance}. We need to decouple the treatment effects from the underlying spatial signal to quantify the effects of the predictors. Second, the exposures of interest in this paper -- e.g., unemployment rate, pay gap, etc. are continuous and hence necessitates the application of the \emph{generalized propensity score} (GPS)  \citep{hirano2004, giffin2023generalized} framework. This utilizes the extension of binary propensity-based causal inference methods for the continuous exposure regimes.

For causal inference, we employ a unified framework combining a regularized spatial basis regression model and a doubly robust GPS estimation. Spatial confounding is addressed by augmenting the regression with low-dimensional spectral basis functions derived from the county adjacency graph -- specifically Moran Eigenvector Maps (MEM) and conditional autoregression (CAR/ICAR)-based basis functions \citep{cressie2022basis, hooten2014simultaneous}. The basis function coefficients are regularized via Lasso regression during model estimation, ensuring that the target treatment effect remains unpenalized to preserve its causal interpretability and eliminate regularization bias. The GPS is then embedded in a doubly robust estimator \citep{kennedy2017non}, which is known to remain consistent if either the propensity or outcome model is correctly specified. Together, this approach aligns with the growing practice of (generalized) propensity score methods for causal analysis of spatially structured observational data\citep{shiba2021using, reich2021review}.

The remainder of the paper is organized as follows. Section~\ref{sec:data_lit}
describes the data and reviews relevant methodological background. Section~\ref{sec:methods} presents the spatial causal inference methodology, including basis construction, GPS estimation, doubly robust estimation, and cross-fitting. Section~\ref{sec:results} reports empirical results on basis function performance and causal effect estimates. Finally, Section~\ref{sec:discussion} discusses implications, limitations, and directions for future work.

\section{Data and Literature Review} \label{sec:data_lit}

\begin{figure}[!ht]
    \centering
    \includegraphics[width=\textwidth]{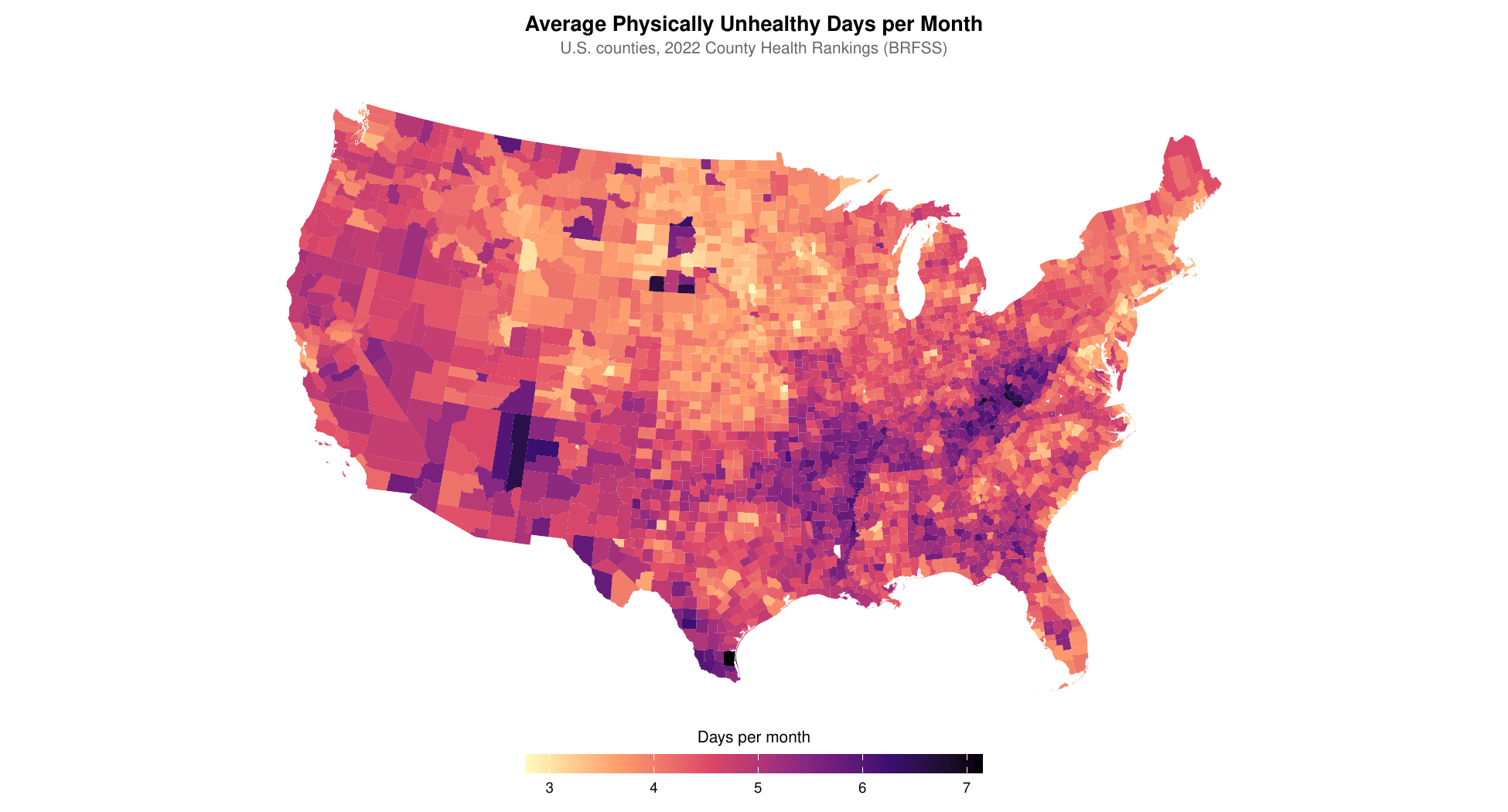}
    \caption{Geographic distribution of average physically unhealthy days per 
    month across U.S. counties (2022 County Health Rankings, derived from BRFSS). 
    Darker shading indicates higher morbidity burden. }
    \label{fig:map_outcome}
\end{figure}

\begin{figure}[ht]
    \centering
    \includegraphics[width=\textwidth]{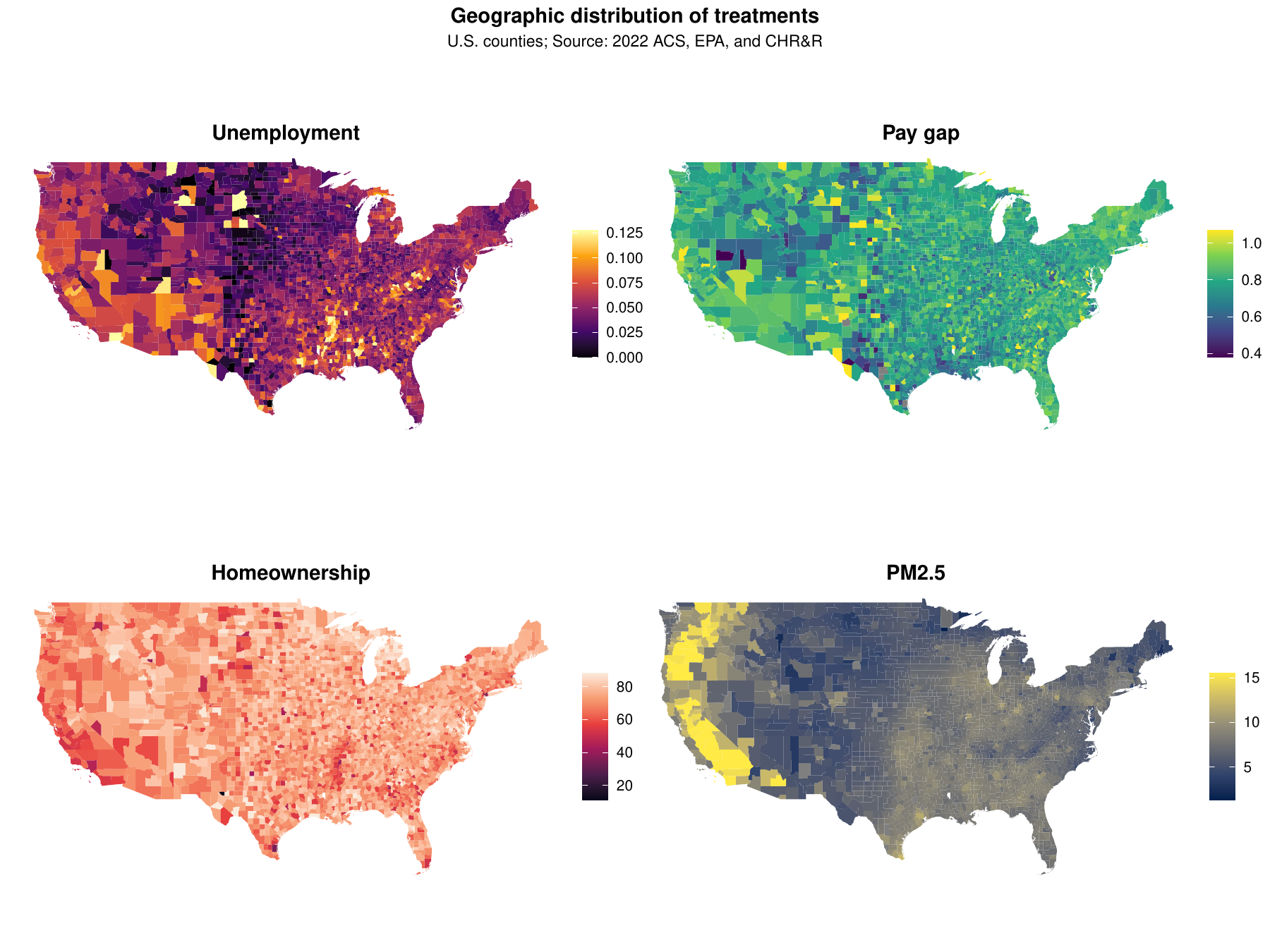}
    \caption{Geographic distribution of the four socioeconomic and environmental 
    treatments across U.S. counties. Top-left shows unemployment rate, top-right shows pay gap, bottom left shows 
    homeownership rate,  and bottom right shows  the PM\,2.5 concentration (EPA via CHR\&R). All the plots are winsorized to cap the extreme 1\% high values.}
    \label{fig:map_treatments}
\end{figure}

Our analysis draws on county-level data over the U.S. counties and 
county-equivalents. The outcome variable -- average physically unhealthy days per 
month -- comes from the 2022 County Health Rankings \& Roadmaps (CHR\&R), derived 
from the CDC Behavioral Risk Factor Surveillance System (BRFSS), and has been 
validated as a reliable proxy for population-level functional limitation and chronic  disease burden \citep{remington2015county}. Figure \ref{fig:map_outcome} shows the spatial distribution of the response variable, revealing a strong geographic pattern in physically unhealthy days across counties.

\begin{figure}
    \centering
    \includegraphics[width=0.5\linewidth, height = 6cm]{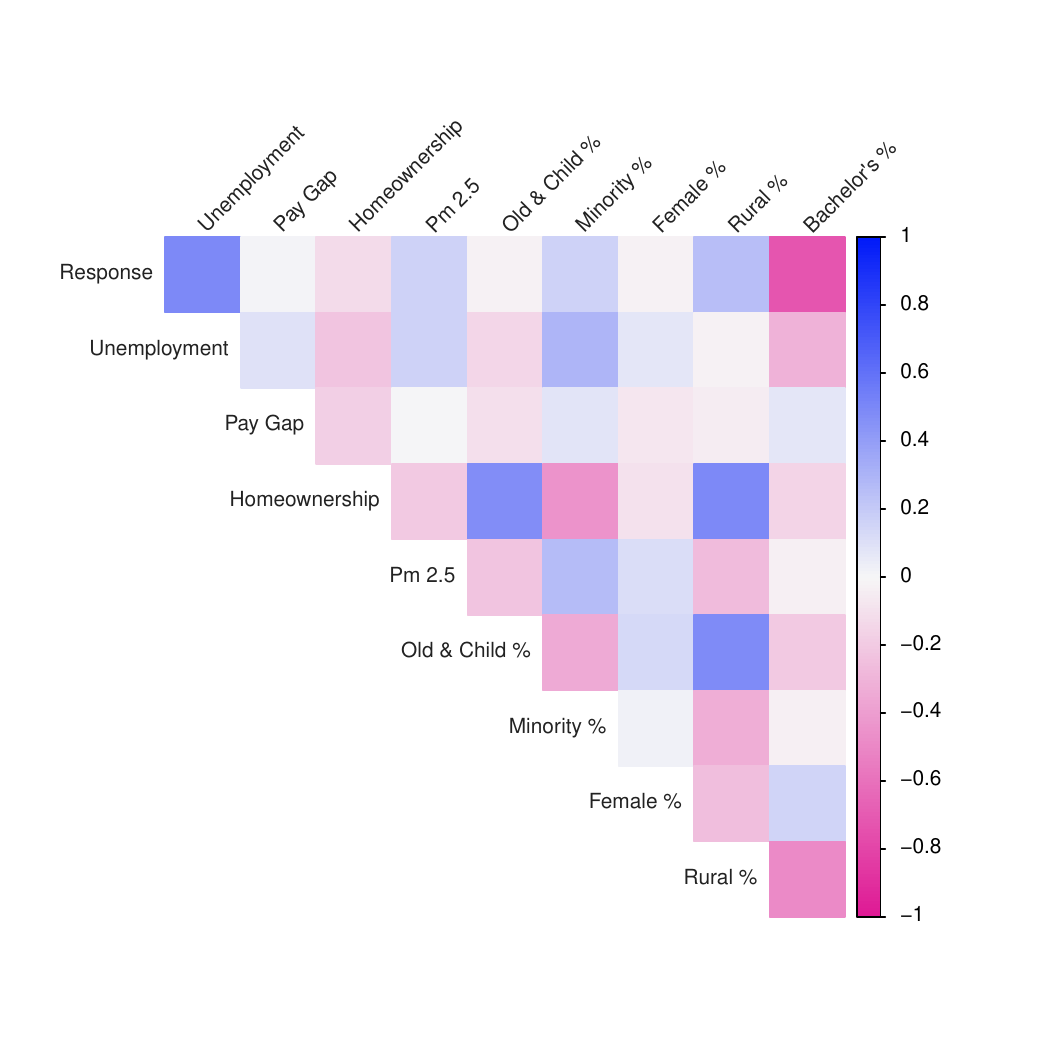}
    \caption{Correlation between the response variable (monthly physically unhealthy days), treatments, and confounder variables from the 2022 CHR\&R data.}
    \label{fig:correlation}
\end{figure}

Predictor data are obtained in an analysis-ready format from the same CHR\&R source, as well as from the American Community Survey (ACS) via \texttt{data.census.gov}. Figure~\ref{fig:map_treatments} shows the spatial distribution of the four treatment variables considered in this study. Here, we are interested in the effect of four treatments -- unemployment rate, 
pay gap, homeownership rate, and PM\,2.5 concentration on the average physically unhealthy days. In addition, we use the percentage of population aged 65+ and under 18 (denoted as Old \& Children \%), percentage of minority population (black, hispanic, and asian), percentage female, percentage rural,  and percentage of bachelor’s degree holders as confounders that may affect both the outcome and the treatment variables. The pairwise correlation matrix of the variables is presented in Figure \ref{fig:correlation}. All variables are measured at the county level, which serves as the unit of analysis for constructing the spatial model through the adjacency graph underlying our basis functions. Related social determinant of health  analyses in the literature include cross-sectional rankings \citep{remington2015county}, structural determinants frameworks \citep{braveman2011social}, geographically weighted regression \citep{rivera2022spatial}, and Bayesian spatiotemporal models 
\citep{comas2025self} -- though most of this work remains associative rather than 
causal.

\textcolor{black}{Our goal is to estimate the causal effects of the drivers on the average number of physically unhealthy days. As noted above, this requires accounting for both the confounders and the spatial dependence structure among all variables \citep{hodges2010adding, hanks2015restricted, paciorek2010importance}. We address these issues using a generalized propensity score-adjusted spatial basis function regression framework. The spatial dependence component is handled through a low-rank basis function approach, in which the full-rank spatial covariance estimation problem is replaced by a low-rank functional representation \citep{cressie2008fixed, cressie2022basis, wikle2010low}. Specifically, we consider two spectral bases in this setting, derived from two graph-based operators on the neighborhood adjacency graph: Moran Eigenvector Maps (MEM) \citep{griffith2014spatial, dray2006spatial} and the \mbox{(I)CAR} precision matrix \citep{besag1991bayesian, hughes2013dimension}. For causal estimation with observational data, we use inverse probability weighting, which for continuous predictors leads to the generalized propensity score (GPS) framework \citep{hirano2004, robins2000marginal, hernan2010causal}, embedded within a cross-fitted Double Machine Learning (DML) doubly robust estimator \citep{robins2001comment, funk2011doubly, kennedy2017non, chernozhukov2018double}. Related recent work at this intersection includes \citet{gao2022causal}, \citet{papadogeorgou2023spatial}, and \citet{pollmann2020causal}, among others.}

\textcolor{black}{It is worth situating our approach relative to related spatial-causal frameworks that address neighborhood dependence under the umbrella of \emph{spillover effects} \citep{giffin2023generalized, forastiere2021identification}; see also the references in \citet{pollmann2020causal}. These methods typically relax SUTVA by explicitly modeling a spatial or network lag of the exposure -- a summary of neighboring units' treatments -- as part of the propensity score, targeting both direct and spillover causal effects. \citet{forastiere2021identification} formalize this through the Stable Unit Treatment Neighborhood Value Assumption (SUTNVA), under which potential outcomes depend on a unit's own treatment and a scalar summary of its neighbors' treatments, with identification proceeding via a GPS that jointly balances individual and neighborhood treatment. Our approach, by contrast, is philosophically closer to the \emph{latent ignorability} framework of \citet{reich2021review}: we do not explicitly relax SUTVA via a neighborhood treatment lag, but instead bypass the need to model physical interference by framing the spatial structure as a problem of unmeasured confounding. Under this setup, we absorb the latent spatial signal from neighboring counties through spectral basis functions (MEM or ICAR eigenvectors) that enter both the GPS and the outcome model. Under this representation, standard causal identification holds conditionally given the confounders and the spatial basis functions. Our basis functions capture the spatial structure of unmeasured confounders rather than an explicit function of neighboring treatments, and we do not separately target spillover effects as a distinct estimand. We additionally employ stabilized inverse probability weights \citep{robins2000marginal}, which improves finite-sample efficiency and weight stability relative to the unstabilized weighting in \citet{giffin2023generalized}.}

\section{Methodology} \label{sec:methods}

In this section, we describe our full causal effect estimation framework below: notations  are described in Section~\ref{sec:data-not}, the basis function idea and construction is presented in Section~\ref{sub:basis}, and the doubly robust causal estimation framework is presented in Section~\ref{sub:robust}.
\begin{figure}
    \centering
    \includegraphics[width=0.8\linewidth]{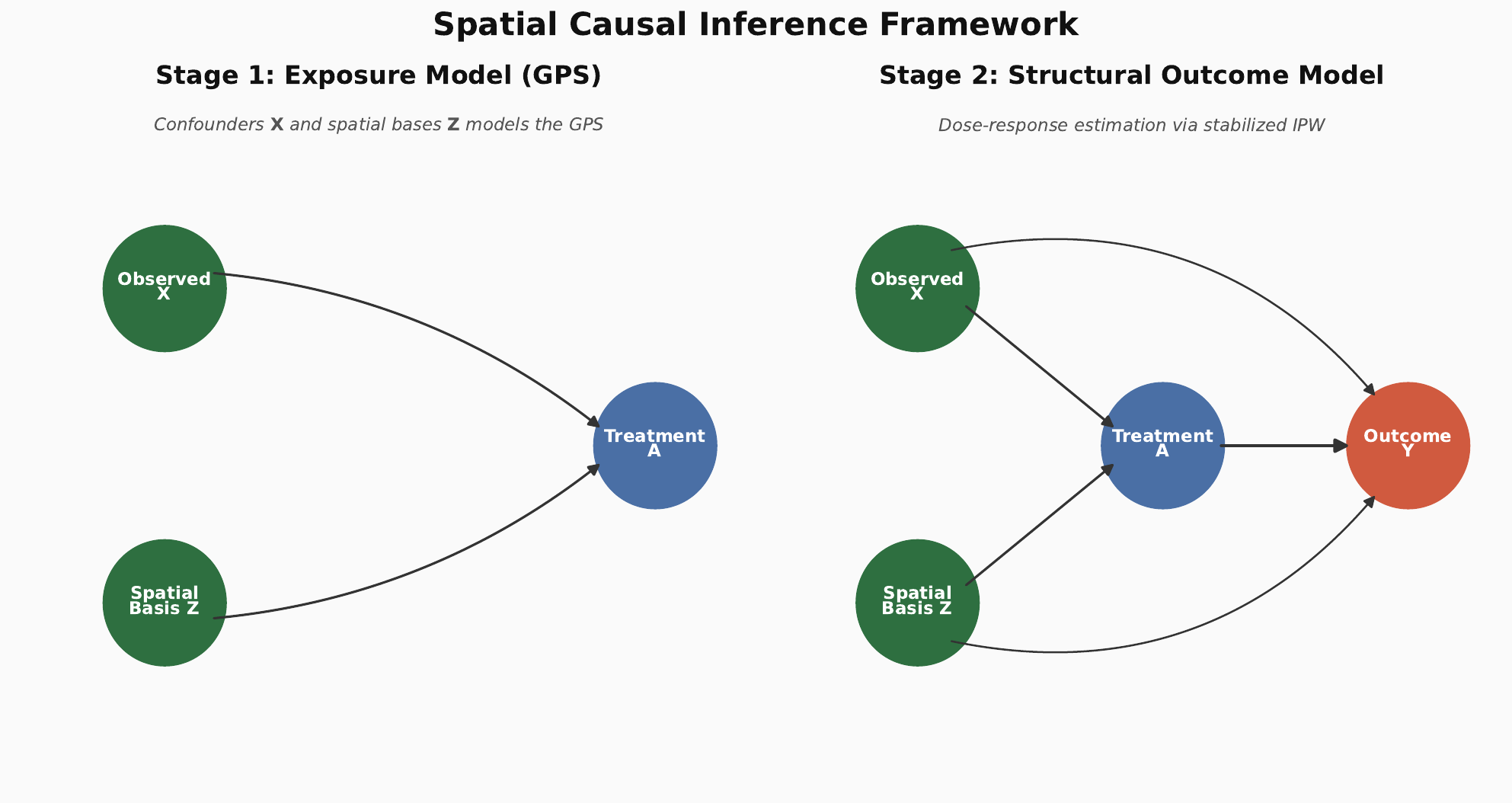}
    \caption{Schematic presentation of the full spatial causal inference framework.}
    \label{fig:pipeline}
\end{figure}
\subsection{Notations} \label{sec:data-not}

Suppose we have data over $n$ counties indexed by $j = 1, \dots, n$. Let $y_j$ denote the response variable (average physically unhealthy days per month) for county $j$. Suppose that we are interested in modeling the effect of $m$ treatment variables denoted by $A_{\ell}$, where $\ell = 1, \dots, m.$ We denote $A_{\ell,j}$ as the value of the treatment variable for county $j.$  Here we separately analyze the causal effect of $m=4$ socioeconomic and environmental treatments -- unemployment rate,
pay gap, homeownership rate, and PM\,2.5 concentration -- on the monthly average physically unhealthy days.

Let $\mathbf {X}_j$ denote a set of additional socioeconomic covariates (confounders), including percentage of population aged 65+ and under 18, percentage of minority population (black, hispanic, and asian), percentage female, percentage rural, and percentage of bachelor’s degree holders. Additionally,  we denote $\Z_j \in \mathbb{R}^{k}$ 
as a set of spatial basis functions for county $j$, which we define in Section~\ref{sub:basis}.

Let $y_j(a_\ell)$ denote the potential outcome corresponding to treatment $A_{\ell} = a_{\ell}; \; \ell=1,\dots,m$. Our objective in this article is to  estimate the average treatment effect $\tau_{\ell} = \mathbb{E}\left[ y_j( A_{\ell})\right]$ of each treatment. We next provide an overview of basis functions as the random effect component for spatial modeling.

\subsection{Spatial Basis Construction} \label{sub:basis}

\color{black}A central connection that motivates our approach is the equivalence between low-rank basis function representations and spatial random effects models, which is well established in the spatial statistics literature \citep{cressie2008fixed,cressie2022basis,wikle2010low}. In the classical spatial mixed model, the relationship between the response, treatment, and confounding variables can be written as
\begin{equation}
    y_j = \mathbf{X}_j^\top \boldsymbol{\gamma} + \beta_{\ell} A_{\ell,j} 
    + u_j + \varepsilon_j, \quad 
    \varepsilon_j \overset{\text{iid}}{\sim} N(0, \sigma^2),
\end{equation}
where $u_j$ is a spatially structured random effect capturing latent geographic variation. Instead of modeling $u_j$ via a full covariance-based approach, basis function representations offer a tractable alternative by linearizing the random effect as
\begin{equation}
    u_j \approx \mathbf{Z}_j^\top \boldsymbol{\alpha} 
    = Z_{j1}\alpha_1 + \cdots + Z_{jK}\alpha_K,
\end{equation}
where each $Z_{jk}$ is a pre-determined function constructed to capture a distinct pattern of spatial variation, and $\boldsymbol{\alpha}$ are the associated coefficients. Once this substitution is made, $u_j$ enters the model as a standard regression term, making the spatial adjustment directly compatible with the causal estimation framework.

There are many available basis function families and modeling strategies. Here we focus on spectral basis functions that provide a low-rank representation of spatial variation. For point-referenced data, these bases are often constructed from the eigendecomposition of a positive definite kernel or covariance operator \citep[see a review in][]{daw2022overview}. The leading eigenfunctions, corresponding to the largest eigenvalues, capture dominant large-scale spatial patterns, while the trailing eigenfunctions represent finer-scale local fluctuations. Retaining only a small number $K$ of leading eigenfunctions yields a low-rank approximation, sometimes called fixed-rank kriging \citep{cressie2008fixed}, with the residual variation absorbed into $\varepsilon_j$ and treated as approximately independent noise.

For the areal data setting considered here, the county adjacency graph motivates analogous graph-based spectral bases. Let $\W \in \{0,1\}^{n \times n}$ denote the binary adjacency matrix, where $w_{ij}=1$ if counties $i$ and $j$ share either a boundary or a vertex (queen contiguity). We consider two such bases, each targeting a different notion of spatial smoothness.

\paragraph{Moran Eigenvector Maps (MEM):}
The MEM basis is motivated by Moran's $I$, a standard measure of spatial autocorrelation in areal data. Eigenvectors that maximize Moran's $I$ represent the most spatially coherent patterns on the graph and thus provide a natural basis for capturing positive spatial dependence. Concretely, the MEM basis is obtained from the doubly centered adjacency matrix
\begin{equation}
    \widetilde{\W} = \left(\I - \frac{\mathbf{11}^\top}{n}\right) 
    \W \left(\I - \frac{\mathbf{11}^\top}{n}\right),
\end{equation}
whose eigenvectors, ordered by decreasing eigenvalue, successively maximize Moran's $I$ and isolate patterns of positive spatial autocorrelation at decreasing spatial scales \citep{griffith2021interpreting, dray2006spatial}. We retain the first $K$ eigenvectors to form $\Z_{\text{MEM}} \in \mathbb{R}^{n \times K}$.

\paragraph{ICAR Basis:}
In contrast, the intrinsic conditional autoregressive (ICAR) model \citep{besag1991bayesian} is defined through conditional dependence on neighboring areas. Under the ICAR prior, the latent spatial effect at each county depends only on its neighbors, leading to the graph Laplacian precision matrix $Q = D - \rho W$, where $D$ is the diagonal degree matrix and $\rho$ is the autocorrelation parameter. In the intrinsic case, $\rho = 1$. Since $Q$ is a precision matrix, the eigenvectors corresponding to its $K$ smallest nonzero eigenvalues span the subspace of large-scale spatial variation \citep{hughes2013dimension} and are used to form $\Z_{\text{ICAR}} \in \mathbb{R}^{n \times K}$.

Here we set $K = 350$ based on the minimum number of basis functions from the set $\mathcal{K} = 50 \times \{1, 2, \dots\}$ that led to no residual spatial autocorrelation using Moran's $I$ test (see Table~\ref{tab:comparison}). Moreover, to avoid overfitting, the regression coefficients on these basis functions are penalized via an $\mathcal{L}_1$ penalty in the causal estimation procedure described in Section~\ref{sub:robust}.

\color{black}

\subsection{Assumptions and Doubly Robust Causal Estimation} \label{sub:robust}

Here we first specify our assumptions required for the causal inference, and then we proceed with the full doubly robust causal framework. For causal effect identification, the following assumptions are required: 
\begin{enumerate}
    \item \textbf{Unconfoundedness (Conditional ignorability):} $y_j(A_{\ell}) \perp\!\!\!\perp A_{\ell, j} \mid \X_j, \Z_j$ for all $A_{\ell, j}$ in the support of $A_{\ell, j}$. Potential outcomes are independent of treatment assignment conditional on confounders and spatial basis. This states there is no unmeasured confounding after conditioning on $\X_j$ and $\Z_j$. \textcolor{black}{This is analogous to the latent ignorability assumption of \citet{reich2021review}, where the spatial basis $\Z_j$ plays the role of the latent spatial confounder.}
    
    \item \textbf{Overlap (Positivity):} The conditional density  $f(A_{\ell, j} \mid \X_j, \Z_j)$ is positive over the relevant support of the confounders. This ensures that all confounder values have non-negligible probability of receiving a range of treatment levels, supporting comparability across treatment values.
    
    \item \textbf{Consistency (Well-defined interventions):} $A_{\ell, j} = a_{\ell} \implies y_j = y_j(a_{\ell})$. This means that the observed outcome under treatment $A_{\ell, j} = a_{\ell}$ equals the potential outcome under the realized treatment $a_{\ell}$. This assumption requires no interference between units (SUTVA) and well-defined treatment effects. \textcolor{black}{Conditional on the spatial basis $\mathbf{Z}_j,$ we treat the residual variation as approximately independent across counties, so that SUTVA holds conditionally given the neighborhood representation encoded in $\mathbf{Z}_j$.}
\end{enumerate}

\noindent Following the assumptions, here we specify the full causal effect estimation framework. Here we employ a doubly robust (DR) estimator based on the generalized propensity score (GPS) for continuous treatments \citep{hirano2004}. A schematic diagram of the DR framework is provided in Figure \ref{fig:pipeline}. It combines two nuisance functions:

\begin{enumerate}[(i)]
    \item \textbf{GPS estimation:} For the $\ell$-th treatment, we model its conditional distribution given all other variables:
    \begin{align}
        A_{\ell, j} \mid \mathbf{X}_j,  \mathbf{Z}_j &\sim N(\mu_{\ell}(\mathbf{X}_j, \mathbf{Z}_j), \sigma_{\ell}^2).\label{modgps}
    \end{align}
    The conditional mean model $\mu_{\ell}$ is estimated via a regularized regression with $A_{\ell, j}$ as the response and the remaining confounder set $(\mathbf{X}_j,  \mathbf{Z}_j)$ as the predictor: $$\mu_{\ell}(\mathbf{X}_j, \mathbf{Z}_j) = \X_j^{\top} \boldsymbol{\gamma}_{\ell} + \Z_j^{\top} \boldsymbol{\alpha}_{\ell}.$$ Here we use an $\mathcal{L}_1$ penalty on spatial basis coefficients. Residual variance $\sigma^2_{\ell}$ is estimated using the residual mean square error. Following that, the estimated GPS is given by the conditional normal density
    \begin{equation}
        f(A_{\ell, j} \mid \mathbf{X}_j,  \mathbf{Z}_j) = \phi(A_{\ell, j}; \widehat{\mu}_{\ell}(\mathbf{X}_j,  \mathbf{Z}_j), \widehat{\sigma}_{\ell}^2). \label{modgps2}
    \end{equation}
    This leads to the construction of stabilized inverse probability weights \citep{robins2000marginal, naimi2014constructing}:
    \begin{equation}
        w_{j\ell} = \frac{f(A_{\ell, j})}{f(A_{\ell, j} \mid \mathbf{X}_j,  \mathbf{Z}_j)}, \label{eqgps}
    \end{equation}
    where $f(A_{\ell, j})$ is the marginal density of the treatment $A_{\ell, j}$. We estimate it using the empirical distribution of $A_{\ell, j}$. This creates a pseudo-population where $A_{\ell, j}$ is balanced across confounders $\mathbf{X}_j$ and $\mathbf{Z}_j$ \citep{robins2000marginal, hernan2010causal}.
    \item \textbf{Outcome model:} For the $\ell$-th treatment, we cross-fit the baseline conditional mean of the outcome using an unweighted regularized lasso regression:
    \begin{equation}
        y_j \mid \mathbf{X}_j, A_{\ell, j}, \mathbf{Z}_j \sim N(g_{\ell,j}(\mathbf{X}_j, A_{\ell, j}, \mathbf{Z}_j), \, \zeta^2_{\ell}), \label{modoutcome}
    \end{equation}
    where $g_{\ell, j}$ represents the conditional mean outcome. To construct predictions $\widehat{g}_{\ell, j}$ avoiding over-fitting bias, spatial basis coefficients are penalized via an $\mathcal{L}_1$ penalty, while coefficients on the confounders and the treatment are left unpenalized.
\end{enumerate}

\noindent Following the stabilization of the generalized propensity score (GPS) weights, the doubly robust estimator for the continuous marginal treatment effect $\tau_{\ell}$ is calculated as a semiparametric additive adjustment to the baseline parametric outcome coefficient:
\begin{equation}
\widehat{\tau}_{\ell} = \widehat{\beta}_{\ell} + \frac{\frac{1}{n}\sum_{j=1}^n w_{j\ell} (A_{\ell, j} - \widehat{\mu}_{\ell,j})(y_j - \widehat{g}_{\ell, j})}{\frac{1}{n}\sum_{i=1}^n (A_{\ell, i} - \widehat{\mu}_{\ell,i})A_{\ell, i}}, \label{eqdr}
\end{equation}
where $\widehat{\beta}_{\ell}$ is the unpenalized treatment coefficient extracted from the full-sample outcome model. The residual variance parameter ${\zeta}^2_{\ell}$ is estimated using the empirical residual mean square error of this regression. This formulation balances the continuous assignment distribution across confounders, ensuring consistency if \textit{either} the GPS mean model or the outcome model is correctly specified. 

Standard errors are computed via the sample variance of the empirical influence function ($\widehat{\phi}_j$), which accounts for both the adjustment score and the variance profile of the baseline parametric component:
\begin{equation}
    \text{SE}(\widehat{\tau}_{\ell}) = \frac{1}{n}\sqrt{ \sum_{j=1}^n \widehat{\phi}_j^2},
\end{equation}
where
\begin{equation}
    \widehat{\phi}_j = \frac{w_{j\ell}(A_{\ell, j} - \widehat{\mu}_{\ell,j})(y_j - \widehat{g}_{\ell,j})}{\frac{1}{n}\sum_{i=1}^n (A_{\ell, i} - \widehat{\mu}_{\ell,i})A_{\ell, i}} + \frac{(A_{\ell, j} - \widehat{\mu}_{\ell,j})A_{\ell,j}}{\frac{1}{n}\sum_{i=1}^n (A_{\ell, i} - \widehat{\mu}_{\ell,i})A_{\ell, i}}(\widehat{\beta}_{\ell} - \widehat{\tau}_{\ell}).
\end{equation}

\color{black}
\noindent We implement $k$-fold cross-fitting to control overfitting bias in the doubly robust estimator. Specifically, we partition the data into $10$ mutually exclusive folds $\{\mathcal{I}_k\},k=1 \dots, 10$. For each fold $k$, we treat $\mathcal{I}_k$ as the held-out (test) set and use the remaining data $\mathcal{I}_k^c$ as the training set. The procedure is as follows:
\begin{enumerate}
    \item \textbf{GPS estimation:} For each fold $k$, using only the training data $\mathcal{I}_k^c$, estimate the GPS by fitting a Lasso regression with $A_{\ell, j}$ as the response and $(\mathbf{X}_j, \mathbf{Z}_j)$ as predictors. The penalty parameter $\lambda_{\text{GPS},\ell}$ is selected via cross-validation on the training data. We then obtain cross-fitted predictions $\widehat{\mu}_{\ell, j}$ for all $j \in \mathcal{I}_k$. Repeating this over all folds yields $\widehat{\mu}_{\ell, j}$ for all observations.

    \item \textbf{Outcome model:} Similarly, for each fold $k$, we fit the outcome model using $\mathcal{I}_k^c$ and obtain predictions on $\mathcal{I}_k$. The penalty parameter $\lambda_{\text{Outcome},\ell}$ is selected via cross-validation. For interpretability, only the spatial basis coefficients $\mathbf{Z}_j$ are penalized in this step, while treatment and confounder effects remain unpenalized. This yields cross-fitted outcome predictions $\widehat{g}_{\ell, j}$ for all $j$.

    \item \textbf{DR Estimator:} Using the cross-fitted estimates $\{\widehat{\mu}_{j\ell}, \widehat{g}_{j\ell}\}_{j=1}^n$, we compute the doubly robust estimator as defined in Equation~\eqref{eqdr}.

\end{enumerate}

\section{Results} \label{sec:results}

\noindent Table \ref{tab:comparison} compares MEM and ICAR across basis dimensions $K$ $\in$ $\{100,$ $150,\, 200, \dots, 500\}$. ICAR marginally but consistently outperforms MEM on predictive accuracy across all values of $K$, and is therefore used in the causal estimation that follows. Both bases successfully eliminate residual spatial autocorrelation by $K = 350$ (Moran's $I = 0.004$, $p = 0.34$ for ICAR), with prediction performance remaining stable beyond that point. We adopt $K = 350$ as the smallest basis dimension that achieves adequate spatial deconfounding.

\begin{table}[htbp]
\centering
\small 
\begin{tabular}{c l r r r c r}
    \toprule
    Number of Bases ($K$) & Basis Type & RMSE ($10^{-1}$) & MAE ($10^{-1}$) & $R^2$ & Active Bases Count & Moran's $p$-val \\ 
    \midrule \hline
    100 & MEM &   2.74 & 2.10 & 0.83 &  93 & 0.00 \\ 
       & ICAR &  2.62 & 2.01 & 0.84 &  96 & 0.00 \\ 
   150 & MEM &   2.67 & 2.04 & 0.83 & 131 & 0.00 \\ 
       & ICAR &  2.53 & 1.93 & 0.85 & 145 & 0.00 \\ 
   200 & MEM &   2.65 & 2.02 & 0.84 & 169 & 0.00 \\ 
       & ICAR &  2.52 & 1.92 & 0.85 & 180 & 0.00 \\ 
   250 & MEM &   2.62 & 2.01 & 0.84 & 203 & 0.00 \\ 
       & ICAR &  2.53 & 1.92 & 0.85 & 208 & 0.00 \\ 
   300 & MEM &   2.62 & 2.01 & 0.84 & 231 & 0.00 \\ 
       & ICAR &  2.52 & 1.91 & 0.85 & 238 & 0.01 \\ 
   350 & MEM &   2.60 & 1.99 & 0.84 & 263 & 0.35 \\ 
       & ICAR &  2.50 & 1.89 & 0.86 & 283 & 0.34 \\ 
   400 & MEM &   2.59 & 1.99 & 0.84 & 293 & 0.83 \\ 
       & ICAR &  2.50 & 1.89 & 0.85 & 298 & 0.62 \\ 
   450 & MEM &   2.58 & 1.99 & 0.85 & 321 & 0.99 \\ 
       & ICAR &  2.51 & 1.89 & 0.85 & 328 & 0.89 \\ 
   500 & MEM &   2.58 & 1.99 & 0.85 & 356 & 1.00 \\ 
       & ICAR &  2.52 & 1.91 & 0.85 & 340 & 0.90 \\ 
   \hline
\end{tabular}
\caption{Comparative performance of MEM and ICAR spatial bases across varying dimensions ($K$). Note the transition to non-significant Moran's $I$ results ($p > 0.05$) at $K \ge 350$, indicating successful removal of residual spatial autocorrelation.}
\label{tab:comparison}
\end{table}

Table~\ref{tab:causal_effects} reports doubly robust GPS estimates for four socioeconomic and environmental determinants, all of which pass the residual autocorrelation check (Moran's $I$, $p > 0.05$). Unemployment rate emerges as the dominant treatment out of all four, with a one percentage point increase in the unemployment rate corresponding to approximately $0.0543$ additional unhealthy days per month (or roughly $0.65$ additional days per year). Consequently, a five percentage point increase in a county's unemployment rate translates to an annualized burden of approximately $3.26$ additional physically unhealthy days per capita. Pay gap has a small but significant effect on the response as well. The positive sign of the coefficient implies that counties with a narrower gender pay gap exhibit a higher average number of physically unhealthy days per month. Homeownership is also significant in the negative direction, plausibly reflecting its role as a broad proxy for household economic stability. PM\,2.5 shows no significant effects after spatial adjustment -- a finding that may reflect weak marginal relationships at the county scale, attenuation from geographic aggregation, or both.

\begin{table}[htbp]
\centering
\begin{tabular}{lrrrr}
\hline
Treatment & Effect & SE & Lower 95\% Bound & Upper 95\% Bound \\
\hline
Unemployment Rate &  $5.437$ & $0.662$ & $4.139$ & $6.735$   \\
Pay Gap           &  $0.388$ & $0.085$ & $0.220$ & $0.555$   \\
Homeownership     & $-0.020$ & $0.002$ & $-0.024$ &  $-0.015$  \\
PM\,2.5           &  $-0.004$ & $0.003$ & $-0.011$ & $0.003$ \\
\hline
\end{tabular}
\caption{Doubly robust causal effect estimates of all treatments}
\label{tab:causal_effects}
\end{table}

\section{Discussion} \label{sec:discussion}

This study applied a spatial causal inference framework combining basis function regularization with doubly robust GPS estimation to identify socioeconomic drivers of physical morbidity across U.S. counties. ICAR-derived eigenvectors marginally but consistently outperformed MEM in both prediction accuracy and residual autocorrelation removal. Among the four treatments examined, unemployment rate is the dominant causal driver of physically unhealthy days, with pay gap modestly significant and homeownership borderline negative. PM\,2.5 shows no significant effects after spatial adjustment, plausibly reflecting attenuation from county-level aggregation or mediation through other pathways.

\textcolor{black}{To discuss the limitations of our approach, we note that some confounding variables are naturally correlated in socioeconomic observational data, which may induce multicollinearity. We address this through Lasso-penalized GPS and outcome models, which provide regularization and implicit variable selection; however, penalization alone does not replace a well-justified causal graph, and our results should be interpreted accordingly. Moreover, the causal ordering of some variables cannot be fully justified. For instance, one may argue that higher unemployment could influence educational attainment. Such directional assumptions are untestable from the observed data. Future work could formalize the causal ordering through expert elicitation and address multicollinearity more explicitly through simulation and methodological extensions.}

Several additional limitations warrant consideration. The data are cross-sectional and county-level, limiting causal direction and introducing ecological fallacy concerns only based on the current version of the data. Basis dimension selection also presents a fundamental problem: standard information criteria such as Akaike Information Criterion (AIC), Bayesian Information Criterion (BIC), Deviance Information Criterion (DIC) and residual autocorrelation diagnostics do not usually agree here. Our chosen $K = 350$ prioritizes the elimination of spatial dependence over model simplicity, and we acknowledge that a full-rank ICAR or Gaussian Process models could provide a more theoretically complete (albeit computationally intensive) alternative. It should be noted though that prediction performance has remained competitive throughout.

Future research should extend this framework into a spatiotemporal formulation. Utilizing multiple releases of CHR\&R data would allow for the exploitation of within-county variation over time, specifically to test the stability of unemployment effects across major shocks like the COVID-19 pandemic. Additionally, more health-related outcomes such as mentally unhealthy days, or even a multivariate framework modeling physical and mental health outcomes jointly would likely yield a more overall understanding of how structural disadvantage shapes the American health landscape.

\backmatter

\section*{Declarations}

The authors report no external funding and no competing interests. All data used in this study are publicly 
available from the 2022 County Health Rankings 
\& Roadmaps (\url{https://www.countyhealthrankings.org}) and the U.S. Census Bureau American Community Survey 5-year estimates \url{https://data.census.gov}.  All authors contributed equally to this 
work.

\bibliography{sn-bibliography}

\newpage 

\color{black}
\begin{appendices}
\section{Consistency and Double Robustness of the Estimator}

We consider the partially linear model (e.g., \citet{chernozhukov2018double})
\[
y_j = \tau_\ell A_{\ell,j} + g_\ell(\mathbf{X}_j, \mathbf{Z}_j) + \varepsilon_j, 
\quad \mathbb{E}[\varepsilon_j \mid A_{\ell,j}, \mathbf{X}_j, \mathbf{Z}_j] = 0,
\]
where \(g_\ell(\mathbf{X}_j, \mathbf{Z}_j)\) captures the effect of observed and spatial confounders. Let \(\mu_{\ell,j} = \mathbb{E}[A_{\ell,j} \mid \mathbf{X}_j, \mathbf{Z}_j]\), and let \(w_{j\ell}\) represent the stabilized inverse probability weights. Following cross-fitting, the doubly robust estimator for the continuous marginal treatment effect \(\tau_{\ell}\) is defined as:
\begin{equation}
\widehat{\tau}_{\ell}
=
\widehat{\beta}_{\ell}
+
\frac{\frac{1}{n}\sum_{j=1}^n w_{j\ell} (A_{\ell,j} - \widehat{\mu}_{\ell,j})(y_j - \widehat{g}_{\ell,j})}
{\frac{1}{n}\sum_{j=1}^n (A_{\ell,j} - \widehat{\mu}_{\ell,j})A_{\ell,j}},
\label{eqdr_proof}
\end{equation}
where \(\widehat{\beta}_\ell\) is the unpenalized treatment coefficient from the full-sample outcome model, \(\widehat{\mu}_{\ell,j}\) is the cross-fitted conditional treatment mean prediction, and \(\widehat{g}_{\ell,j}\) is the cross-fitted outcome expectation, similar to \citet{robins1994estimation}.

\paragraph{Proof.}

Let \(\delta_{g,j} = \widehat{g}_{\ell,j} - g_{\ell,j}\), \(\delta_{\mu,j} = \widehat{\mu}_{\ell,j} - \mu_{\ell,j}\), and define the treatment residual as \(\widetilde{A}_{\ell,j} = A_{\ell,j} - \widehat{\mu}_{\ell,j}\). Substituting the structural model for \(y_j\), the outcome residual decomposes as:
\[
y_j - \widehat{g}_{\ell,j} = \tau_\ell A_{\ell,j} - \delta_{g,j} + \varepsilon_j.
\]

\noindent We define the empirical adjustment correction term \(\widehat{C}_\ell\) as:
\[
\widehat{C}_\ell =
\frac{\frac{1}{n}\sum_{j=1}^n w_{j\ell}\widetilde{A}_{\ell,j}(y_j - \widehat{g}_{\ell,j})}
{\frac{1}{n}\sum_{i=1}^n \widetilde{A}_{\ell,i}A_{\ell,i}}.
\]

\noindent Substituting the outcome residual decomposition yields:
\[
\widehat{C}_\ell = T_{1n} + T_{2n} + T_{3n},
\]
where
\[
T_{1n} =
\tau_\ell \cdot
\frac{\frac{1}{n}\sum_{j=1}^n w_{j\ell}\widetilde{A}_{\ell,j}A_{\ell,j}}
{\frac{1}{n}\sum_{i=1}^n \widetilde{A}_{\ell,i}A_{\ell,i}},
\quad
T_{2n} =
-\frac{\frac{1}{n}\sum_{j=1}^n w_{j\ell}\widetilde{A}_{\ell,j}\delta_{g,j}}
{\frac{1}{n}\sum_{i=1}^n \widetilde{A}_{\ell,i}A_{\ell,i}},
\quad
T_{3n} =
\frac{\frac{1}{n}\sum_{j=1}^n w_{j\ell}\widetilde{A}_{\ell,j}\varepsilon_j}
{\frac{1}{n}\sum_{i=1}^n \widetilde{A}_{\ell,i}A_{\ell,i}}.
\]

\paragraph{Step 1: Evaluation of the denominator.}
By expanding the empirical instrument in the denominator, we have:
\[
\frac{1}{n}\sum_{i=1}^n \widetilde{A}_{\ell,i}A_{\ell,i}
=
\frac{1}{n}\sum_{i=1}^n (A_{\ell,i} - \mu_{\ell,i} - \delta_{\mu,i})A_{\ell,i}
=
\frac{1}{n}\sum_{i=1}^n (A_{\ell,i} - \mu_{\ell,i})A_{\ell,i}
-
\frac{1}{n}\sum_{i=1}^n \delta_{\mu,i}A_{\ell,i}.
\]
Under cross-fitting and regularity conditions, \(\|\delta_{\mu}\|_2 = o_p(1)\). By the law of large numbers, the first term converges to its expectation:
\[
\frac{1}{n}\sum_{i=1}^n (A_{\ell,i} - \mu_{\ell,i})A_{\ell,i}
\xrightarrow{p}
\mathbb{E}\!\left[(A_{\ell} - \mu_{\ell})A_{\ell}\right]
=
\mathbb{E}\!\left[(A_{\ell} - \mu_{\ell})^2\right]
=
\sigma_\ell^2.
\]
Thus,
\[
\frac{1}{n}\sum_{i=1}^n \widetilde{A}_{\ell,i}A_{\ell,i}
\xrightarrow{p}
\sigma_\ell^2.
\]

\paragraph{Step 2: The leading term (\(T_{1n}\)).}
Now examine the numerator of \(T_{1n}\). Expanding \(\widetilde{A}_{\ell,j}\) gives:
\[
\frac{1}{n}\sum_{j=1}^n w_{j\ell}\widetilde{A}_{\ell,j}A_{\ell,j}
=
\frac{1}{n}\sum_{j=1}^n w_{j\ell}(A_{\ell,j} - \mu_{\ell,j})A_{\ell,j}
-
\frac{1}{n}\sum_{j=1}^n w_{j\ell}\delta_{\mu,j}A_{\ell,j}.
\]
The second term is \(o_p(1)\). For the first term, the stabilized weights reweight the sample toward the target pseudo-population, so under correct specification of the generalized propensity score model,
\[
\mathbb{E}\!\left[w_{j\ell}(A_{\ell,j} - \mu_{\ell,j})A_{\ell,j}\right]
=
\mathbb{E}\!\left[(A_{\ell} - \mu_{\ell})A_{\ell}\right]
=
\sigma_\ell^2.
\]
By Slutsky's theorem, the ratio converges to
\[
T_{1n} \xrightarrow{p} \tau_{\ell} \cdot \frac{\sigma_\ell^2}{\sigma_\ell^2} = \tau_{\ell}.
\]

\paragraph{Step 3: Double robustness (\(T_{2n}\)).}
We expand the numerator of \(T_{2n}\) as:
\[
\frac{1}{n}\sum_{j=1}^n w_{j\ell}\widetilde{A}_{\ell,j}\delta_{g,j}
=
\frac{1}{n}\sum_{j=1}^n w_{j\ell}(A_{\ell,j} - \mu_{\ell,j})\delta_{g,j}
-
\frac{1}{n}\sum_{j=1}^n w_{j\ell}\delta_{\mu,j}\delta_{g,j}.
\]
The second term is a product of cross-fitted nuisance estimation errors and is $O_p(\|\delta_{\mu}\|_2 \|\delta_g\|_2) = o_p(n^{-1/2})$ under standard rate conditions. For the first term, two cases arise:
\begin{itemize}
\item \textbf{Case A (correct GPS mean model):} If the propensity mean model \(\mu_{\ell,j}\) is correctly specified, then $\mathbb{E}[A_{\ell,j} - \mu_{\ell,j} \mid \mathbf{X}_j, \mathbf{Z}_j] = 0,$
and the weighted product has mean zero.
\item \textbf{Case B (correct outcome model):} If the baseline outcome expectation \(g_{\ell,j}\) is correctly specified, then \(\delta_{g,j} \xrightarrow{p} 0\), so the term vanishes.
\end{itemize}
Thus, \(T_{2n} = o_p(1)\) if either nuisance model is correctly specified.

\paragraph{Step 4: Sampling variability term (\(T_{3n}\)).}
For \(T_{3n}\), the structural disturbance satisfies $\mathbb{E}[\varepsilon_j \mid A_{\ell,j}, \mathbf{X}_j, \mathbf{Z}_j] = 0.$ Because cross-fitting uses separate folds for nuisance estimation, the summands are mean zero conditional on the estimated nuisance functions. Hence,
\[
\frac{1}{n}\sum_{j=1}^n w_{j\ell}\widetilde{A}_{\ell,j}\varepsilon_j = O_p(n^{-1/2}),
\]
and, since the denominator converges to the positive constant \(\sigma_\ell^2\), it follows that $T_{3n} = O_p(n^{-1/2}) = o_p(1).$

\paragraph{Step 5: Conclusion.}
Combining the three terms yields: $\widehat{C}_\ell = \tau_\ell + o_p(1).$
Therefore, the debiased estimator is consistent:
$$\widehat{\tau}_\ell = \widehat{\beta}_\ell + \widehat{C}_\ell \xrightarrow{p} \tau_\ell.$$
Moreover, consistency holds if either the outcome regression \(g_\ell\) or the treatment model \(\mu_\ell\) is correctly specified, establishing the double robustness property.

\hfill \(\square\)

\section{Additional Results}

Table~\ref{tab:coeffs} reports the estimated coefficients for the confounding variables from the outcome models across the four treatment specifications.

\begin{table}[]
    \centering
    \begin{tabular}{|l|c|c|c|c|c|}
    \hline
       Treatment  & Old \& Child \% & Minority \% & Female \% & Rural \% & Bachelor \%  \\
       \hline
       Unemployment Rate & -1.071 & 0.003 & 0.020 & 0.001 & -5.700\\
       Pay Gap & -1.295 & 0.004 & 0.028 & 0.001 & -6.417 \\
       Homeownership & 0.058 & 0.001 & 0.014 & 0.003 & -6.221\\
       PM\,2.5  &  -1.339 & 0.005 & 0.026 & 0.001 & -6.386\\
       \hline
    \end{tabular}
    \caption{Estimated coefficients for confounders from the final outcome models under each treatment specification.}
    \label{tab:coeffs}
\end{table}

\begin{figure}
    \centering
    \subfloat[]{\includegraphics[width=0.48\textwidth]{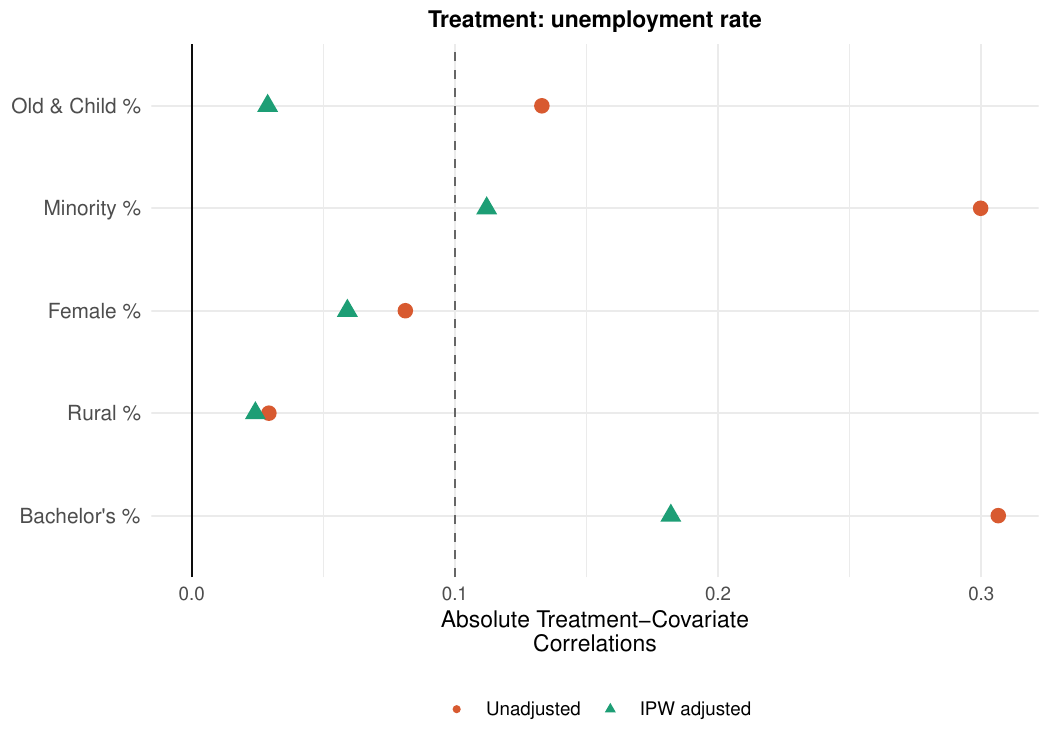}} \hfill
    \subfloat[]{\includegraphics[width=0.48\textwidth]{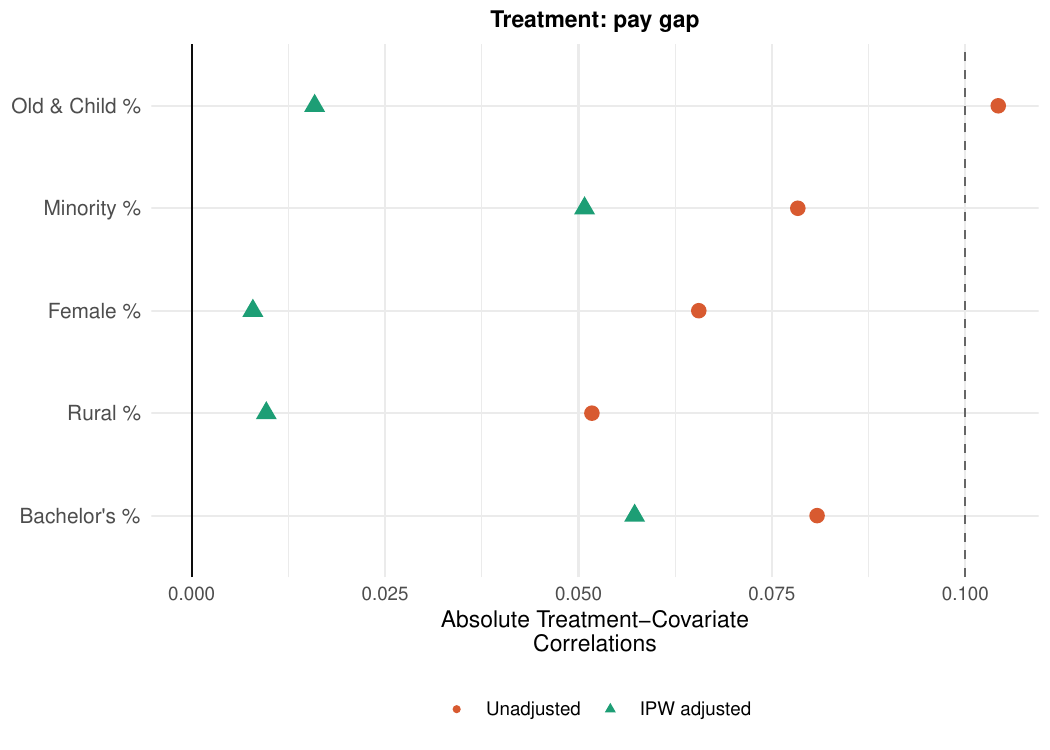}} \\ \vspace{0.2cm} 
    \subfloat[]{\includegraphics[width=0.48\textwidth]{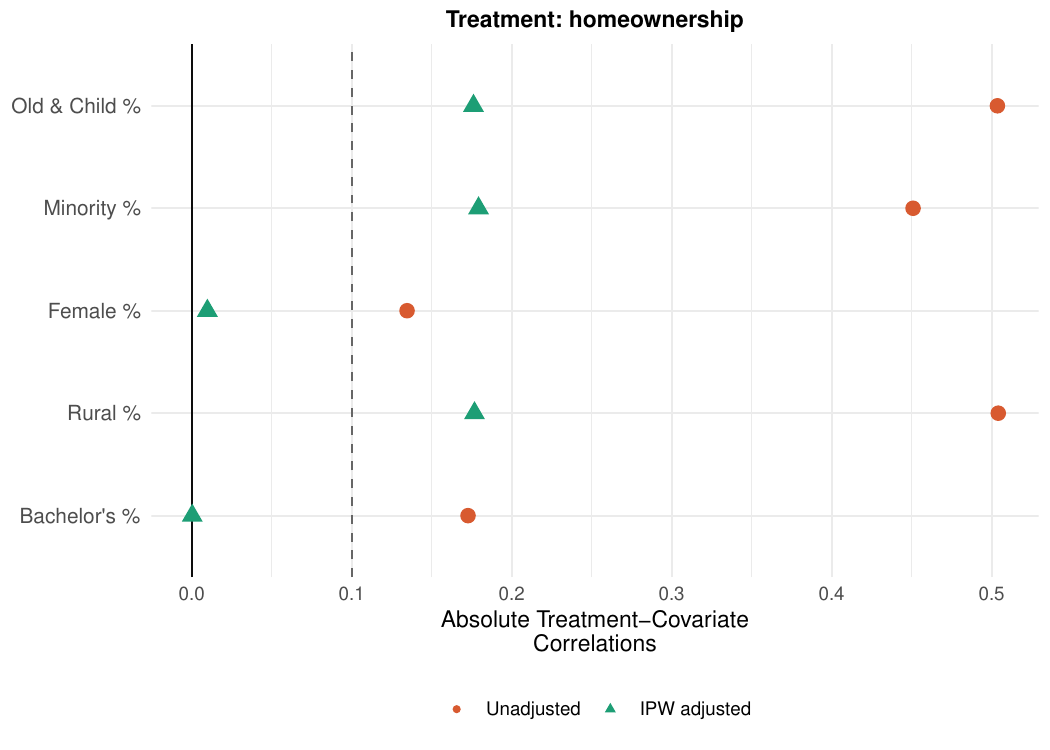}} \hfill
    \subfloat[]{\includegraphics[width=0.48\textwidth]{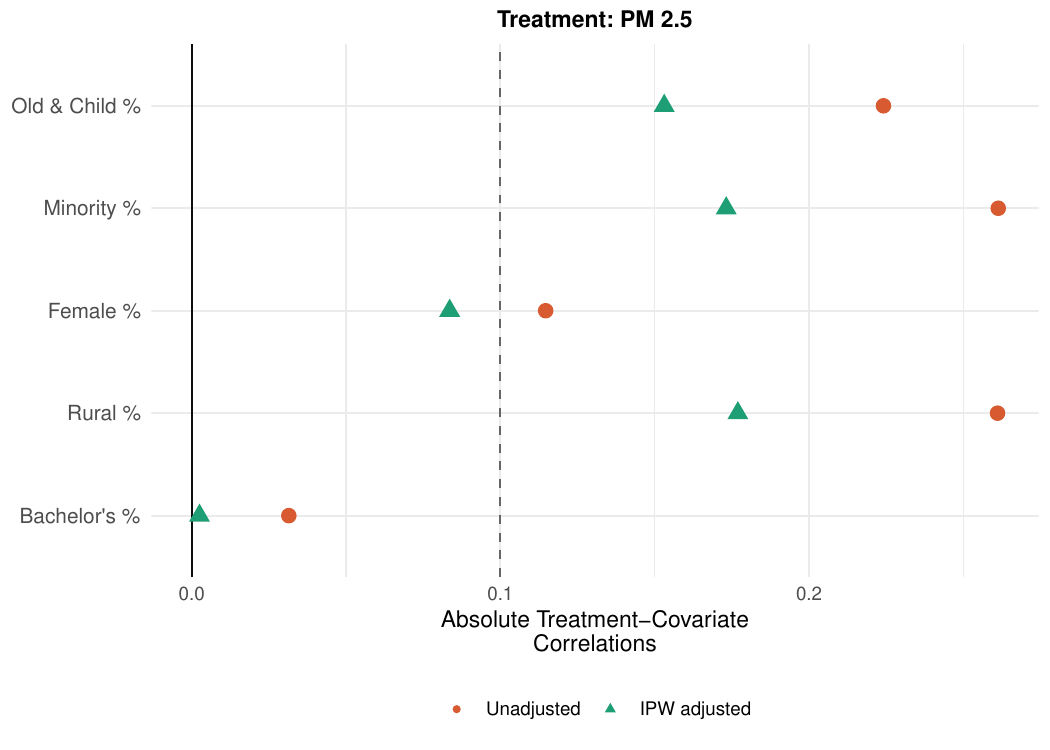}}
    \caption{Covariate balance diagnostics for the four treatment specifications: (a) unemployment rate, (b) pay gap, (c) homeownership, and (d) PM\,2.5.}
    \label{fig:loveplot}
\end{figure}

Figure~\ref{fig:loveplot} presents the corresponding covariate balance diagnostics (love plots) for the four treatment models. The results indicate that the generalized propensity score (GPS) adjustment substantially reduces the correlation between the treatments and the observed confounders. While a small number of covariates remain above the conventional threshold of 0.1, which is not uncommon in complex observational settings, the overall reduction in imbalance is consistent across all specifications.

\end{appendices}
\end{document}